\documentstyle[aps,preprint,epsf]{revtex}

\def\beginwide{
        \end{multicols} \vspace*{-0.5cm} \noindent
        \rule{3.5in}{.1mm}\rule{.1mm}{5mm} \widetext \medskip }
\def\beginwidetop{
        \end{multicols} \vspace*{-0.5cm} \noindent
        \widetext \medskip }
\def\endwide{
        \hspace*{3.35in}~\rule[-5mm]{.1mm}{5mm}\rule{3.5in}{.1mm}
        \begin{multicols}{2} \vspace*{-1.0cm} \noindent }
\def\endwidebottom{
        \begin{multicols}{2} \vspace*{-1.0cm} \noindent }
\draft
\begin{document}

\title{Low field hysteresis in disordered ferromagnets}

\author{Lorenzo Dante$^1$, Gianfranco Durin$^2$, Alessandro Magni$^2$
and Stefano Zapperi$^1$}

\address{$^1$ INFM unit\`a di Roma 1, Dipartimento di Fisica,
    Universit\`a "La Sapienza", P.le A. Moro 2
        00185 Roma, Italy\\
    $^2$ Istituto Elettrotecnico Nazionale Galileo Ferraris and
        INFM, strada delle Cacce 91, I-10135 Torino, Italy
        }
\maketitle
\begin{abstract}
We analyze low field hysteresis close to the demagnetized
state in disordered ferromagnets using the zero temperature
random-field Ising model. We solve the demagnetization process
exactly in one dimension and derive the Rayleigh law of hysteresis.
The initial susceptibility $a$ and the hysteretic coefficient $b$
display a peak as a function of the disorder width. This behavior
is confirmed by numerical simulations $d=2,3$ showing that in
limit of weak disorder demagnetization is not
possible and the Rayleigh law is not defined.
These results are in agreement with experimental observations
on nanocrystalline magnetic materials.
\end{abstract}


\section{Introduction}
Ferromagnetic materials display hysteresis under the action of an external field and the
magnetization depends in a complex way on the field history. In order to define magnetic
properties unambiguously, it is customary to first {\it demagnetize} the material,
bringing it to a state of zero magnetization at zero field. This can be done, in
practice, by the application of a slowly varying AC field with decreasing amplitude. In
this way, the system explores a complex energy landscape, due to the interplay between
structural disorder and interactions, until it is trapped into a low energy minimum. This
demagnetized state is then used as a reference frame to characterize the magnetic
properties of the material.

The hysteresis properties at low fields, starting from the demagnetized state, have been
investigated already in 1887 by Lord Rayleigh \cite{RAY-87}, who found that the branches
of the hysteresis loop are well described by parabolas. In particular, when the field is
cycled between $\pm H^*$ , the magnetization $M$ follows $M = (a+bH^*)H\pm
b((H^*)^2-H^2)/2$, where the signs $\pm$ distinguish the upper and lower branch of the
loop. Consequently the area of the loop scales with the peak field $H^*$ as $W=4/3 b
(H^*)^3$ and the response to a small field change, starting from the demagnetized state
is given by $M^*=a(H^*)\pm b (H^*)^2$ \cite{Bertotti}.

The Rayleigh law has been widely observed in ferromagnetic materials
\cite{Bertotti}, but also in ferroelectric ceramics \cite{DAM-97,BOL-00}. 
The current theoretical interpretation of this law
is based on a 1942 paper by N\'eel \cite{NEE-42}, who derived the law formulating the
magnetization process as the dynamics of a point (i.e. the position of a domain wall) in
a random potential. In this framework, the initial susceptibility $a$ is associated to
reversible motions inside one of the many minima of the random potential, while the
hysteretic coefficient $b$ is due to irreversible jumps between different valleys.
Successive developments and improvements have been devoted to establish precise links
between N\'eel random potential and the material microstructure
\cite{PFE-67,VER-74,KRO-92,MAG-99}, but in several cases the issue is still unsettled.
For instance, the initial permeability of nanocrystalline materials
typically displays a peak as a function of the grain size \cite{HER-97}, heat treatment
\cite{SUZ-91,LEU-97} or alloy composition \cite{HER-97,LIM-93}. This behavior can be
associated to changes in the disordered microstructure, but can not be accounted for by
N\'eel theory that predicts a monotonic dependence of $a$ on the disorder width
\cite{NEE-42}.

The zero temperature random-field Ising model (RFIM) has been recently used to describe
the competition between quenched disorder and exchange interactions and their effect on
the hysteresis loop \cite{SET-93}. In three and higher dimensions, the model shows a
phase transition between a continuous cycle for strong disorder and a discontinuous loop,
with a macroscopic jump, at low disorder. The two phases are separated by a second order
critical point, characterized by universal scaling laws \cite{SET-93,DAH-96,PER-99} . A
behavior of this kind is not restricted to the RFIM but has also been observed in other
models, with random bonds or random anisotropies \cite{VIV-94} and vectorial
spins \cite{DAS-99}. In addition, a similar disorder induced phase transition in the
hysteresis loop has been experimentally reported for a Co-Co0 bilayer \cite{BER-00}. Thus
the RFIM provides a tractable model for a more generic behavior: the model has been
solved exactly in one dimension \cite{SHU-96,SHU-00} and on the Bethe lattice
\cite{DHA-97,SHU-01}, while mean-field theory \cite{SET-93} and renormalization
group \cite{DAH-96} have been used to analyze the transition.

Here, we use the RFIM to analyze the demagnetization process and investigate the
properties of the hysteresis loop at low fields. Along the lines of
Refs.~\cite{SHU-00,SHU-01}, we compute the demagnetization cycles exactly in one dimension and
derive the Rayleigh law, obtaining $a$ and $b$ as function of disorder and exchange
energies. Next, we analyze the problem numerically in higher dimensions (i.e. $d=2$ and
$d=3$) where exact results are at present not available. In $d=3$, we find that the disorder
induced transition \cite{SET-93}, defined on the saturation loop, is also reflected by
the Rayleigh loops: in the weak disorder phase  the system can not be demagnetized, as
the final magnetization coincides with the saturation magnetization. A similar
behavior has been recently obtained analyzing subloops \cite{CAR-01}. In the high
disorder phase, however, a demagnetization process is possible and hysteresis loops are still described by the Rayleigh law.
Above the transition, the dependence of 
$a$ and $b$ on disorder is qualitatively similar
in all dimensions, displaying a peak and decreasing 
to zero for very strong disorder in
agreement with experiments \cite{HER-97,SUZ-91,LEU-97,LIM-93}.

\section{The random field Ising model}
In the RFIM, a spin $s_i = \pm 1$ is assigned to each
 site $i$ of a $d-$dimensional
lattice. The spins are coupled to their nearest-neighbors 
spins by a ferromagnetic interaction of strength $J$ and 
to the external field $H$. In addition, to each site of
the lattice it is associated a random field $h_i$ taken from given probability
distribution $\rho(h)$. In the following we will mainly focus on a Gaussian
with variance $R$ (i.e. $\rho(h)=\exp(-h^2/2R^2)/\sqrt{2\pi}R$), but we will
also consider a  rectangular distribution.
The Hamiltonian thus reads
\begin{equation} {\cal H} = -\sum_{\langle i,j \rangle}Js_i s_j -\sum_i(H+h_i)s_i,
\label{eq:rfim}
\end{equation}
where the first sum is restricted to nearest-neighbors pairs.
The dynamics proposed in Ref.~\cite{BER-90} and used in Refs.~\cite{SET-93,DAH-96,PER-99}
is such that  the spins align with the local field
\begin{equation}
s_i = \mbox{sign}(J\sum_j s_j  + h_i +H).
\end{equation}

In $d=1$, a spin with $n$ neighbors {\em up}  ($n=0,1,2$), 
will be  {\em up} at the field $H$ with probability: 
\begin{equation}
p_n(H)\equiv \int_{2(1-n)J-H}^{+\infty}\rho(h_i)d\,h_i\end{equation}
When a spin flips {\em up}  the  local field of its neighbors 
is raised by $2J$ so that it can happen
that one or both of the  two neighbors flip {\em up}.
In this way a single spin flip can lead the neighboring spins
to flip, eventually triggering an avalanche.

It has been shown that the RFIM obeys return-point memory \cite{SET-93}: 
if the field is increased adiabatically the magnetization only depends 
on the state in which the field
was last reversed.  This property has been exploited in $d=1$ 
and in the Bethe lattice
to obtain exactly the saturation cycle and the first 
minor loops \cite{SHU-00}. In the next section we will briefly recall
the results reported in Ref.~\cite{SHU-00} and we will then proceed
with a general derivation for nested minor loops.

\section{Saturation loop and first return curves}

\begin{figure}[h]

        \epsfxsize=8cm
        \epsfbox{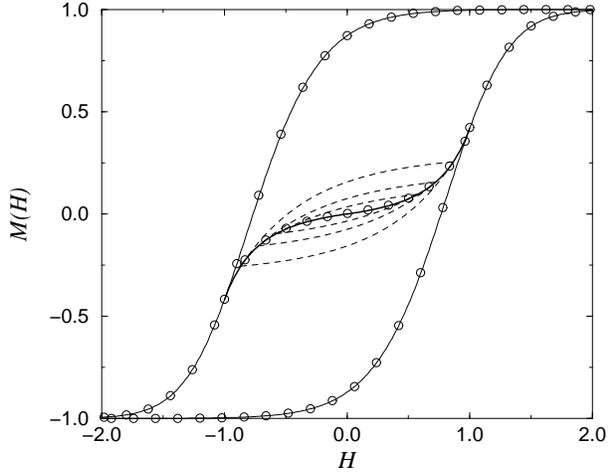}
        \vspace*{1cm}

\caption{Exact expressions for the saturation cycle (thin lines),
the demagnetization curve (thick lines) and a few minor loops
(dotted lines) for $J=1$ and $R=1$. The points are the results
of a numerical simulation with $L=5~10^5$ spins and
a single realization of the disorder.}
\label{fig:1}
\end{figure}

To obtain the saturation loop, we start from the initial condition 
$s_i = -1$ at $H = -\infty$  and we will raise the field up to $H_0$. 
We are thus moving on the lower half of the major hysteresis loop.
Following Ref.~\cite{SHU-00}, we define the conditional probability $U_0$  
that a spin  flips {\em up} at $H_0$ before a given nearest neighbor.
To compute $U_0$, we take advantage of the translational invariance of
the system. There are only two ways  to flip {\em up} 
a spin in  $i$ keeping the spin in $i-1$ {\em down}. 
The two contributions yield $U_0=p_1(H_0)U_0+p_0(H_0)[1-U_0]$, 
from which we obtain 
\begin{equation}
U_0=\frac{p_0(H_0)}{1-\big[p_1(H_0)-p_0(H_0)\big]}
\end{equation}
The probability that a spin  is {\em up} at field $H_0$ is  
\begin{equation}
p(H_0)= U_0^2p_2(H_0)+2U_0(1-U_0)p_1(H_0)+(1-U_0)^2p_0(H_0)
\end{equation}
and the magnetization per spin $M(H_0)$ is simply $M(H_0)=2p(H_0)-1$. 
In Fig.$\ref{fig:1}$ we show the saturation loop for a Gaussian distribution 
of random fields. 

If the field is reversed from a finite value $H_0$, we have a new situation 
and the system departs from the saturation curve. 
It is possible to show that if the field changes from $H_0$ to $H_1=H_0-2J$ 
the magnetization reaches the upper saturation loop again.
Thus we can restrict the analysis to fields included in $[H_0-2J, H_0] $.
The first return curve can be obtained counting the spins 
that were {\em up} at $H_0$ and are {\em down} at $H_1$. 
To this end, we introduce $D_1$ as the conditional probability 
that a spin is {\em down} if its neighbor is {\em up}. Following
similar steps as for $U_0$ \cite{SHU-00}, we obtain: 
\begin{equation}
D_1=\frac{f(H_0)+U_0\left[p_2(H_0)-p_2(H_1)\right]}
{1-\big[p_1(H_0)-p_1(H_1)\big]},
\end{equation}
where $f(H_0)\equiv U_0[1-p_1(H_0)]+(1-U_0)[1-p_2(H_0)]$. 
At this point it is straightforward to write the probability $p(H_1)$  
that a spin is {\em up} at $H_1$:
\begin{eqnarray}
p(H_1)=p(H_0)-
U_0^2\big[p_2(H_0)-p_2(H_1)\big]\\
+2U_0D_1\big[p_1(H_0)-p_1(H_1)\big] 
+D_1^2\big[p_0(H_0)-p_0(H_1)\big]\nonumber
\end{eqnarray}
which is simply related to the magnetization.

\section{Demagnetization}
 
Here, we extend the approach of Ref.~\cite{SHU-00} 
to  more general field histories, treating explicitly 
the demagnetization process: 
the external field is changed through a nested succession $H =
H_0 \to H_1 \to H_2 \to ..... H_n...\to 0$, 
with  
$H_{2n}>H_{2n+2}>0$, $ H_{2n-1}<H_{2n+1}<0$ and
$dH\equiv H_{2n}-H_{2n+2}\to 0$.
The initial value $H_0$ 
should correspond to complete
saturation, but we discussed above that as long as $H_n\geq J$ 
the magnetization $M_n\equiv M(H_n)$  simply follows the saturation curve, 
so that we can set $H_0=J$.

As in the previous section, the key quantity to compute is 
the conditional probability $U_{2n}$ that a spin flips up before 
its nearest neighbor when the field is increased
from $H_{2n-1}$ to $H_{2n}$. Similarly on the descending part of the loops 
we define $D_{2n+1}$ as the conditional probability that a spin flips
 down before its nearest neighbor when the field
is decreased from $H_{2n}$ to $H_{2n+1}$. 
Enumerating all possible spin histories, 
we find recursion relations for the conditional probabilities which 
read as \cite{nota1} 
\begin{equation}\label{eq:u}
\left \{ \begin{array}{l}
 U_{2n} \ \ \ \     =   U_{2n-2}+\left[\displaystyle\frac{U_{2n-2}\big[p_1(H_{2n})\  \,- \ p_1(H_{2n-2})\big]
+ \,D_{2n-1}\big[p_0(H_{2n})  -p_0(H_{2n-2})\big]}{1-\big[p_1(H_{2n}) -p_1(H_{2n-1})\big]}\right]
\\
\\
D_{2n+1}= D_{2n-1}+\left[\displaystyle\frac{D_{2n-1}\big[p_1(H_{2n-1})-p_1(H_{2n+1})\big]
+U_{2n}\big[p_2(H_{2n-1})-p_2(H_{2n+1})\big]}{1-\big[p_1(H_{2n})-p_1(H_{2n+1})\big]}\right].
\end{array}
\right.
\end{equation}
The derivation of Eqs.~(\ref{eq:u})  
is a little involved and we thus report it in the Appendix.

The magnetization as a function of the peak field is
given by
\begin{eqnarray} \label{eq:dem}
M_{2n}=M_{2n-1}+2U_{2n}^2[p_2(H_{2n})-p_2(H_{2n-1})]        \\
+4U_{2n}D_{2n-1}[p_1(H_{2n})-p_1(H_{2n-1})] \nonumber\\
+2D_{2n-1}^2[p_0(H_{2n})-p_0(H_{2n-1})]\nonumber
\end{eqnarray}
and a similar expression holds for $M_{2n+1}$.

In the limit $H_{2n-2}-H_{2n}\equiv dH \to 0$, $H_{2n}\to H^*$
and $H_{2n-1}\to -H^*$, the recursion relations in Eqs.~\ref{eq:u}
become a pair of differential equations \cite{nota2}, 
\begin{equation}\label{eq:dif1}
\left \{ \begin{array}{l}\displaystyle
\frac{\partial U}{\partial H^*}=
\Big[\frac{1}{1-\Omega}\Big]\Big[\rho(H^*)\tilde{D}+ 
\rho(2J-H^*)U\Big]  \\
\qquad\\ \displaystyle
\frac{\partial \tilde{D} }{\partial H^*}=
\Big[\frac{1}{1-\Omega}\Big]\Big[\rho(H^*)U- \rho(2J-H^*)\tilde{D} \Big],      
\end{array}
\right.
\end{equation}
where $\Omega\equiv \int_{-H^*}^{H^*}\rho(h')dh'$ and 
$\tilde{D}(H)\equiv D(-H)$. The boundary conditions are given by
 the conditional  probabilities on the saturation loop 
(i.e. $U(J)=\tilde{D}(J)=U_0(J)=1/2$) and 
the solution reads
\begin{equation}\label{eq:u*}
U(H^*)=\tilde{D}(H^*)={1 \over 2} \exp \Big[-\int^{J}_{H^*}
\frac{\rho(h')+\rho(2J-h')}{1-\Omega(h')} dh' \Big]
\end{equation}
Once the conditional probability $U$ is known, it is straightforward
to compute the magnetization as a function of the peak field $H^*$
from Eq.~\ref{eq:dem}, noting that $M(-H^*)=-M(H^*)$.
Inner loops starting from the demagnetization curve 
(i.e. Eq.~\ref{eq:dem}) can also be computed exactly. 
In Fig.~\ref{fig:1} we report the demagnetization curve and a few inner loops
for a system with Gaussian random field distribution with unit variance. 
The analytical results are compared with numerical simulations, 
performed on a lattice with $L=5~10^5$ spins, using
a single realization of the disorder. 
The perfect agreement between the curves confirms
that the magnetization is self-averaging, as assumed throughout the 
calculations.

\section{Rayleigh law}

To analyze low field hysteresis  we
first substitute in Eq.~\ref{eq:dem} $H_{2n}$ and $H_{2n-1}$ 
with $H^*$ and  $-H^*$. If we  start to reverse the field from 
$H_0=J$ and we cycle the field symmetrically around $H^*=0$, the process displays the 
symmetry $M(H^*)=-M(-H^*)$ and $U(H^*)=\tilde{D}(H^*)$. Thus   we can reduce Eq.~\ref{eq:dem} to
\begin{equation}\label{mm1}
M(H^*)=2U^2(H^*)\sum_{k=0}^{1}\big[p_k(H^*)-p_k(-H^*)\big].
\end{equation}
Now we can  expand $M(H^*)$   around $H^*=0$.
In this limit  we have
\begin{equation}\label{mm2}
[p_k(H^*)-p_k(-H^*)]\simeq \left\{\begin{array}{ll}
2H^*\rho(2J) & \textrm{if $k=0,2$}\\
2H^*\rho(0) \ \  & \textrm{if $k=1$}
\end{array} \right.
\end{equation}
and 
\begin{equation}\label{mm3}
U^2(H^*)\simeq U^2(0)\big[1+2H^*\big(\rho(0)+\rho(2J)\big)\big ].
\end{equation}
Collecting Eq.~\ref{mm2} and Eq.~\ref{mm3} in Eq.~\ref{mm1}, we
obtain $M\simeq aH^*+b(H^*)^2$ recovering the Rayleigh expression
with 

\begin{equation}
\label{a,b}
\left\{\begin{array}{l}
a=4U^2(0)\Big[\rho(0)+\rho(2J)\Big]\\ 
b=4U^2(0)\Big[\rho(0)+\rho(2J)\Big]^2.
\end{array}
\right.
\end{equation}
An expansion can also be performed for minor loops on
the demagnetization curve (i.e cycling $H$ between $\pm H^*$), 
yielding $M = (a+bH^*)H\pm
b((H^*)^2-H^2)/2$, which coincides with the Rayleigh law. 

\begin{figure}[h]
        \epsfxsize=8cm
        \epsfbox{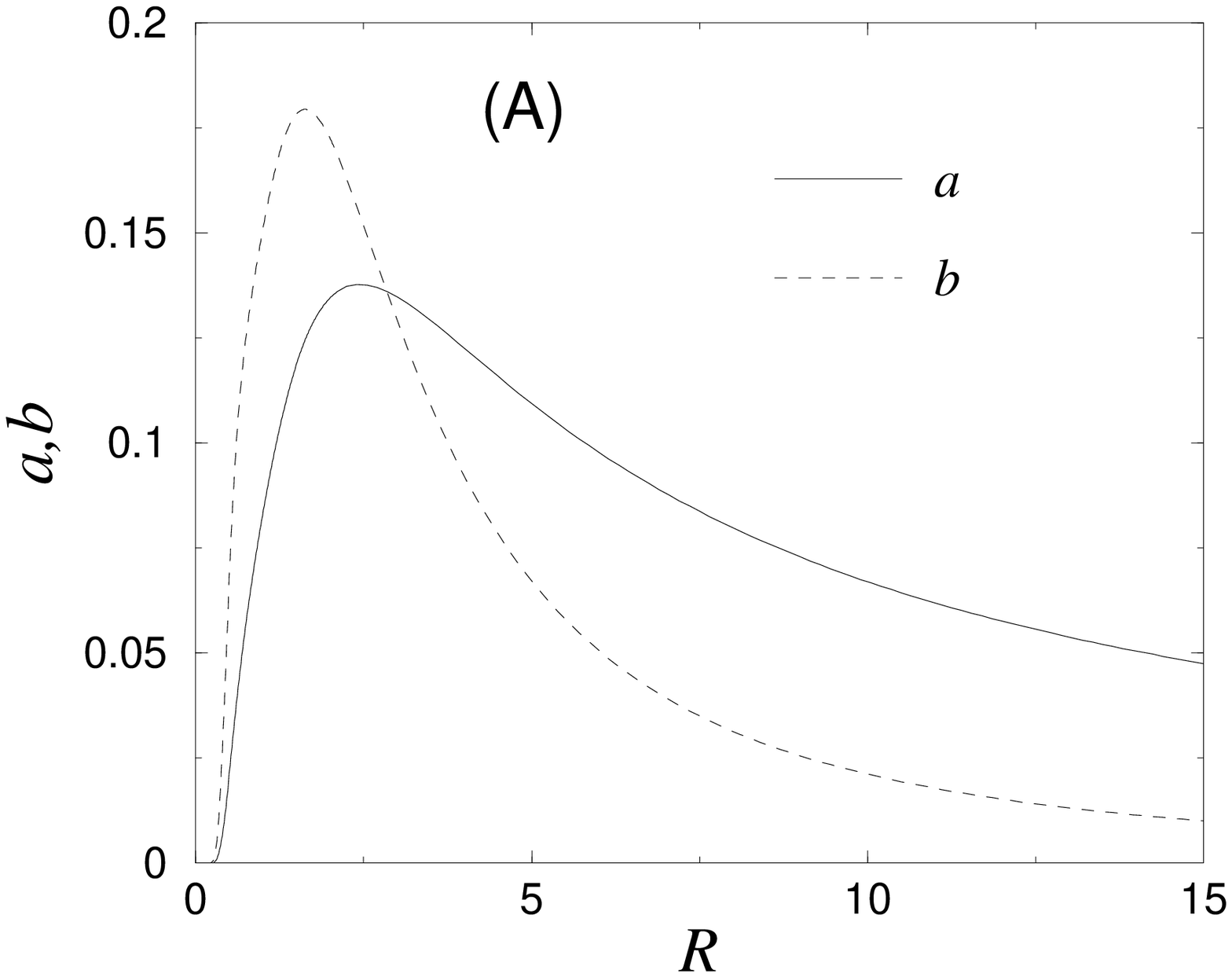}
      	\epsfxsize=8cm
	\epsfbox{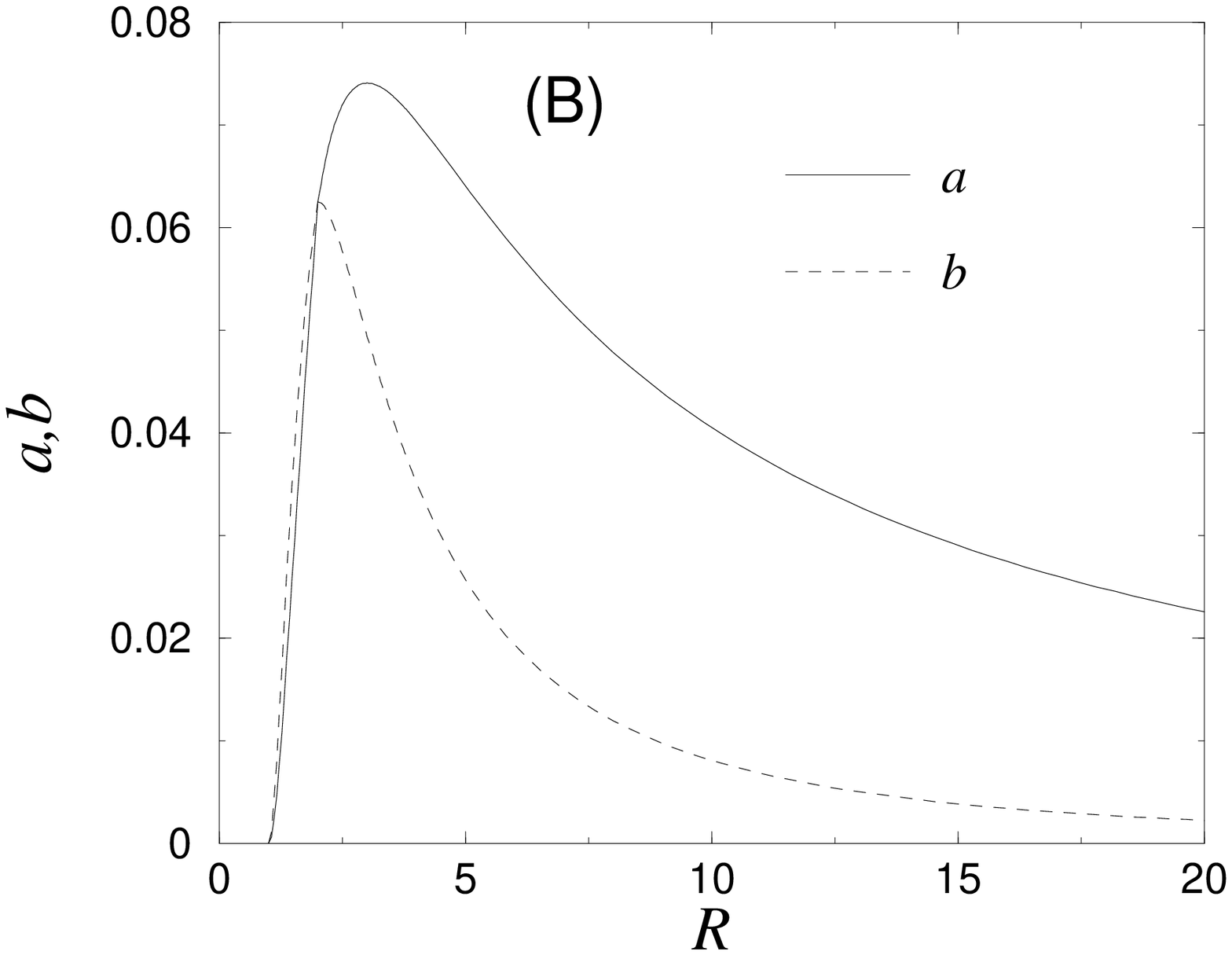}
	\vspace*{1cm}
\caption{The reversible susceptibility $a$ and the 
hysteretic coefficient $b$ computed exactly in
$d=1$ for (a) a Gaussian distribution of random fields  and
(b) a rectangular distribution.}
\label{fig:2}
\end{figure}

In Fig.~\ref{fig:2}a we report the values of $a$ and $b$ for a 
Gaussian distribution of random fields 
as a function of the disorder $R$, showing that both components
of the susceptibility display a maximum in $R$.
To identify the low and strong disorder
behavior of the susceptibilities, we perform an 
asymptotic expansion and we obtain for $R\to \infty$ that
$ a\simeq \displaystyle  \frac{2}{ \sqrt{2\pi}R}    $ and 
$ b\simeq \displaystyle  \frac{2}{ \pi R^2}  $.
For  $ R\to 0$, we obtain:
$a\simeq \displaystyle \bigg(  \frac{1}{e\pi J}\bigg)   e^{-\frac{J^2}{2R^2}}    $ and  
$b\simeq \displaystyle \bigg( \frac{1}{e \pi J\sqrt{2\pi} } \bigg)  \frac{1}{R}e^{-\frac{J^2}{2R^2}}$.
Finally in Fig.~\ref{fig:2}b we report $a$ and $b$ obtained with
a rectangular distribution of random fields. The derivation of these
results is reported in appendix B.

\section{Simulations in $d=2,3$}

Next, we turn our attention to high dimensional system, 
for which analytical results are not available. 
In order to obtain unambiguously the demagnetized state for a given
realization of the disorder, one should perform a 
{\em perfect demagnetization}. This is done in practice changing the 
field by precisely the amount necessary to flip the first
unstable spin. In this way, the field is cycled 
between $-H^*$ and $H^*$ and $H^*$ is then decreased at 
the next cycle by precisely the amount necessary to have
one avalanche less than in the previous cycle. This corresponds to 
decrease $H^*$ at each cycle by an amount $dH$, 
with $dH \to 0^+$. The perfect demagnetization algorithm allows
to obtain a precise characterization of the 
demagnetized state but it is computationally
very demanding. 
Thus we resort to a different algorithm which performs an {\em approximate
demagnetization}: instead of cycling the field between $-H^*$ 
and $H^*$ we just flip the field between these two values and 
then decrease $H^*$ by a fixed amount $dH$. We have
checked that with a reasonably small $dH$ 
(i.e. $dH < 10^{-3}$) the demagnetization curve
is quite insensitive to the algorithm used.

\begin{figure}[h]

        \epsfxsize=8cm
        \epsfbox{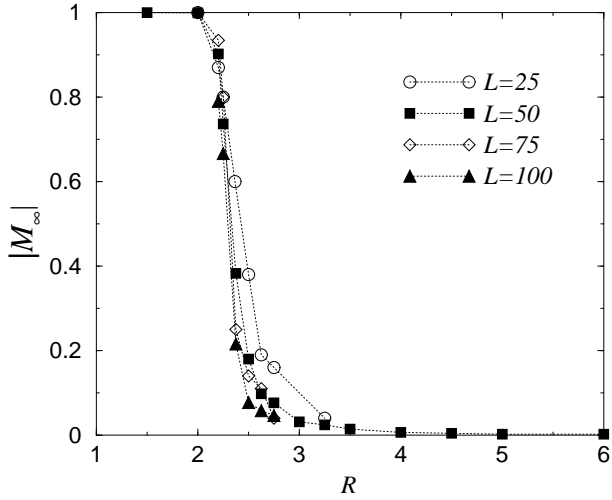}
        \vspace*{1cm}

\caption{The absolute value of the final magnetization $|M_\infty|$ as
a function of $R$, obtained from numerical simulations in $d=3$. For
strong disorder $|M_\infty|=0$ as expected, while for weak disorder the final
magnetization coincides with the saturation value. The transition between
the two types of behavior becomes sharper as the system size is increased.}
\label{fig:3}
\end{figure}

As we discussed above, it is well 
established that in $d=3$ the saturation loops reveal a phase 
transition at $R_c\simeq 2.16$ for $J=1$ \cite{PER-99} 
(the transition is not present in $d=1$, while in $d=2$ the issue is
controversial \cite{PER-99}). We find that the transition 
is reflected also in the Rayleigh loops: in Fig.~\ref{fig:3} 
we report the final magnetization $M_\infty$ computed using the
demagnetization algorithm for different values of $R$. 
For strong disorder $R>R_c$, we see that $M_\infty \simeq 0$ as expected, 
but as $R<R_c$ the demagnetization curve tends to
the saturation magnetization and $M_\infty \to \pm 1$. 
The transition becomes sharper as the system size is increased, 
indicating that demagnetization is possible only for $R>R_c$
(see also Ref.~\cite{CAR-01}). We notice here that two 
scenarios are possible for $L\to \infty$ as $R\to R_c^-$. The first
possibility is that $M_\infty$ scales continuously to zero as
$(R_c-R)^\beta$ and the second is that the transition is
discontinuous (i.e. $M_\infty \to M^* >0$). The present numerical
results do not allow to distinguish between these two cases, but
a recent analysis of the RFIM on the Bethe lattice
is in favor of the first alternative \cite{COL-01}.

\begin{figure}[h]

        \epsfxsize=8cm
        \epsfbox{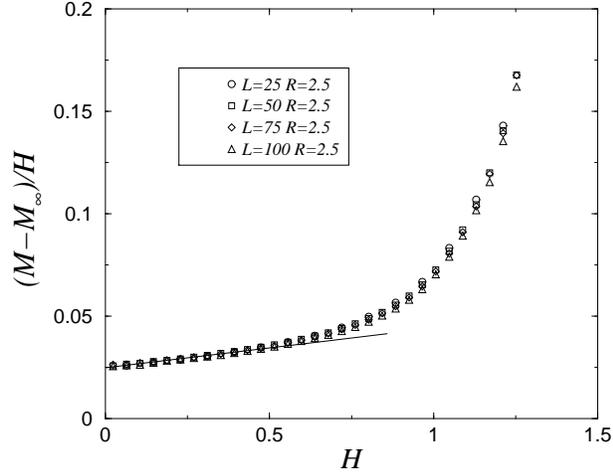}
        \vspace*{1cm}

\caption{The demagnetization curve can be used to obtain
an estimate of the Rayleigh parameters. Notice the absence of
system size dependence. These results are obtained in $d=3$.}
\label{fig:1a}
\end{figure}

\begin{figure}[h]

        \epsfxsize=8cm
        \epsfbox{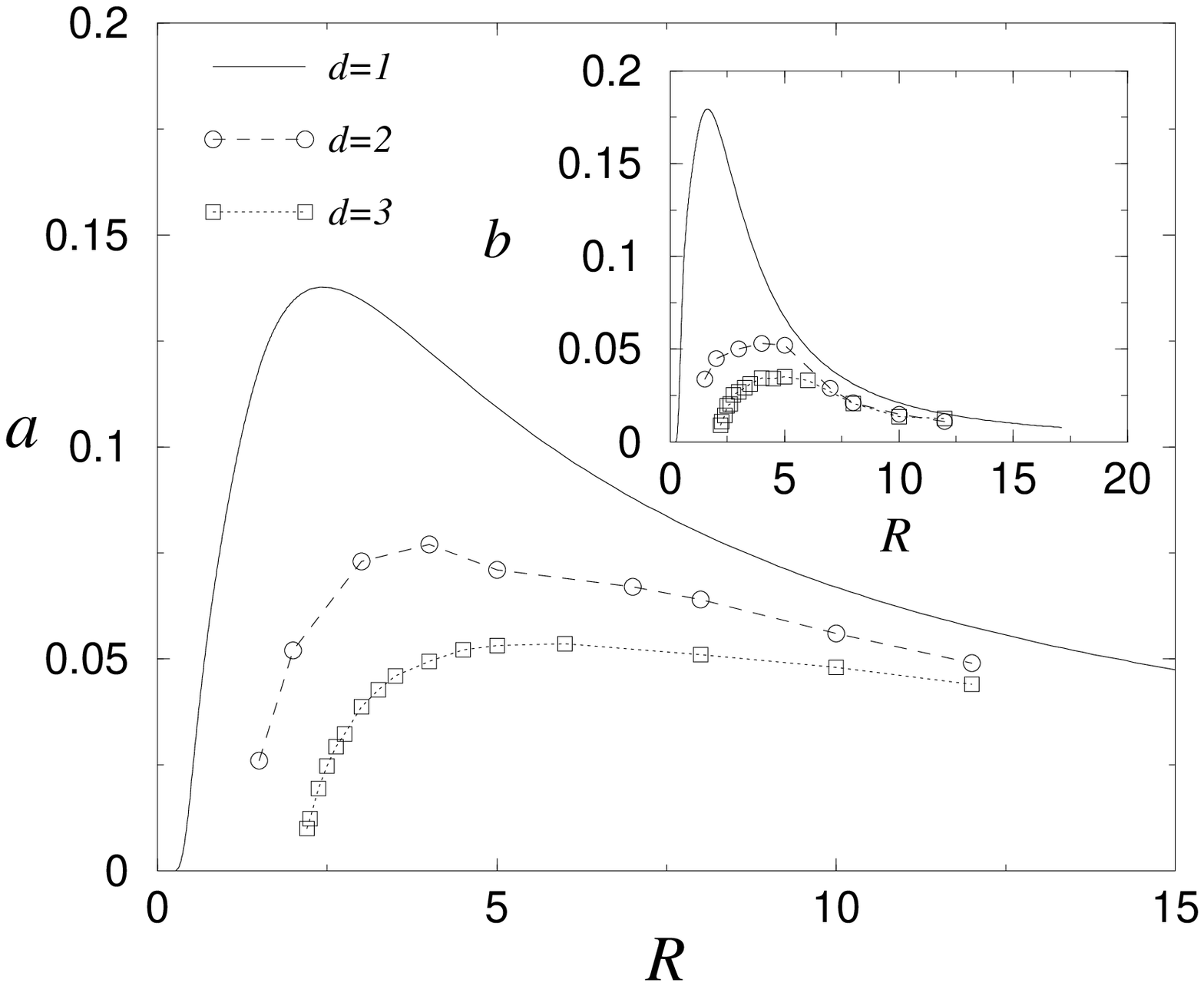}
        \vspace*{1cm}

\caption{The reversible susceptibility $a$ computed exactly in
$d=1$ is compared with numerical results in $d=2$ and $d=3$. In the
inset we show a similar plot for the parameter $b$.}
\label{fig:2b}
\end{figure}

From the demagnetization curve, the Rayleigh parameters
can be estimated plotting $M-M_\infty/H$ vs $H$ and fitting the linear
part of the curve close to $H=0$ (see Fig.~\ref{fig:1a}). As
we show in Fig.~\ref{fig:1a} the demagnetization curve is 
basically independent from the system size, once the 
magnetization has been shifted by $M_\infty$. Thus we
expect that the Rayleigh parameters be also independent
on $L$. In Fig.~\ref{fig:2b} we report the values of $a$ and $b$ 
obtained numerically in $d=2$ and $d=3$ for
different values of $R$, using systems of sizes 
$(L=100)^2$ and $(L=50)^3$. The results
are qualitatively similar to those obtained 
exactly in $d=1$: the curve displays a peak
for intermediate disorder and decrease to zero 
for weak and strong disorder. 

\section{discussion}

In this paper we have discussed the demagnetization properties
of the RFIM in $d=1,2,3$. In $d=1$ it is possible to compute
exactly the demagnetization curve and obtain an expression for
the Rayleigh parameters. We find that $a$ and $b$ display a
peak in the disorder $R$. This result is confirmed by numerical
simulations in $d=2,3$, where analytical results are not available.
In addition, in $d=3$ the disorder induced phase transition 
strongly affects the demagnetization process: for $R<R_c$ it
is not possible to demagnetize the system anymore. 

It is interesting to compare our theoretical results 
with experiments on nanocrystalline
materials. It has been reported that the initial
 susceptibility in several cases displays
a peak as the heat treatment or the alloy composition are varied
\cite{HER-97,SUZ-91,LEU-97,LIM-93}. The peak is usually 
associated to changes in the
microstructure, which induce a competition between the disorder 
present in grain anisotropies and inter-grain interactions mediated by 
the amorphous matrix \cite{HER-97}. Notice that a similar behavior
can not be reproduced by N\'eel theory, where the initial susceptibility
is decreasing with the width of the disorder potential \cite{NEE-42}.
On the other hand, we see here that the behavior is well captured by the 
RFIM, that allows to analyze the the effect of the disorder-exchange 
ratio $R/J$. For weak disorder, we have a few large
domains and the susceptibility is dominated by domain 
wall dynamics. When the disorder is increased, 
the number of domains (and domain walls) also increases and so does the
susceptibility. Increasing the disorder further leads to a 
complete breakup of the domains and the response is dominated 
by single spin flips in low random-field regions
with a progressive decrease of the susceptibility.

A detailed understanding of the demagnetization process and low
field hysteresis has important implications also from 
a purely theoretical point of view. When a disordered 
system is demagnetized, it explores a complex energy landscape
until it finds a metastable minimum. It would be interesting
to compare the statistical properties of the demagnetized state,
with those of the ground state of the system \cite{BER-90}. 
The analysis of the ground state of disordered systems has received a wide 
attention in the past few years, due to the connections with
general optimization problems, and the RFIM is one of the typical
model used to test ground state algorithms \cite{ALA-01}. Demagnetization
could provide a relatively simple way to obtain a low energy state
that can be useful for optimization procedures. 
We are currently pursuing investigations along these lines \cite{ZAR-01}.

\section*{acknowledgments}
This work is supported by the INFM PAIS-G project on ``Hysteresis
in disordered ferromagnets''. We thank M. J. Alava, G. Bertotti, 
F. Colaiori and A. Gabrielli for useful discussions and remarks.

\appendix
\section{Derivation of the recursion relations}

Here, we derive  recursion relations for the conditional probabilities 
$U_{2n}$ and $D_{2n+1}$ as a function of the previous magnetization history.
Let us first consider the case of $D_{2n+1}$:
the field from $H_{2n-1}$  reaches $H_{2n}$ and is then
decreased again up to $H_{2n+1}$.
The weight of the fraction of spins that at field $H_{2n+1}$ flip {\em down} 
before their neighbor is given by
\begin{equation}\label{DDD}
D_{2n+1}=D_{2n-1}-\zeta_{2n}+\zeta_{2n+1},
\end{equation}
where $\zeta_{2n}$ is the weight of the fraction of spins that were 
{\em down} at $H_{2n-1}$ before a fixed nearest neighbor and flip 
{\em up} at  $H_{2n}$, while $\zeta_{2n+1}$ is the weight 
of the fraction of spins contributing to
$\zeta_{2n}$ which  flip again {\em down}  at $H_{2n+1}$.  

To compute  $\zeta_{2n}$, we  consider the spins that at the field 
$H_{2n-1}$ are {\em down} before their
neighbor (for instance, we can say that the spin $i$-th is {\em down} 
before the spin in site $i-1$) and are {\em up} at the field $ H_{2n}$. 
Since we fixed {\em up} the spin in site $i-1$, the spin in site 
$i+1$ can be either  {\em up} or {\em down}.
If the spin in $i+1$ is  {\em up} when the spin 
$i$ flips {\em up}, it contribution to $\zeta_{2n}$ with
\[
U_{2n}\big[p_2(H_{2n})-p_2(H_{2n-1})\big].
\]
If the spin in site $i+1$ is  {\em down} when the spin $i$ flips 
{\em up}, we obtain
\[
D_{2n-1}\big[p_1(h_{2n})-p_1(h_{2n-1})\big].
\]
Indeed, \(\big[p_n(H_{2n})-p_n(H_{2n-1})\big]\) is the probability
 that a spin with $n$ {\em up} nearest neighbors is  {\em up} 
at $H_{2n}$  but not at $H_{2n-1}$, while $D_{2n-1},U_{2n}$ are respectively 
the conditional probabilities that the spin in site $i+1$ is {
\em down} or {\em up} if the spin in site $i$ is {\em down}. 
Adding the two contributions, we obtain 
\begin{equation}
\zeta_{2n}=\Big\{ D_{2n-1}\big[p_1(H_{2n})-p_1(H_{2n-1})\big]+
U_{2n}\big[p_2(H_{2n})-p_2(H_{2n-1})\big]\Big\}.
\end{equation}
The derivation of $\zeta_{2n+1}$ follows similar steps: we count
the spins that are {\em up} at $H_{2n}$ and 
are again {\em down} at $H_{2n+1}$.
If the spin in the site $i+1$ is {\em up} at $H_{2n+1}$, 
the spin in $i$ is {\em up} at $H_{2n}$ and is {\em down} at $H_{2n+1}$ with
probability
\[
U_{2n}\big[p_2(H_{2n})-p_2(H_{2n+1})\big].
\]
Finally, we analyze the case in which the spin in site $i+1$ is already 
{\em down} when the spin $i$ flips {\em down}.
The weight of this configuration is
\[
D_{2n+1}\big[p_1(h_{2n})-p_1(h_{2n+1})\big],
\]
so that $\zeta_{2n+1}$ is given by 
\begin{equation}
\zeta_{2n+1} =\Big\{U_{2n}\big[p_2(H_{2n})-p_2(H_{2n+1})\big]
+D_{2n+1}\big[p_1(H_{2n})-p_1(H_{2n+1})\big]\Big\}.
\end{equation}
Substitute these two expressions in Eq.~(\ref{DDD})
we obtain the second of Eqs.~\ref{eq:u}.
We can then derive a similar equation for $U_{2n}$ 
(First of Eqs.~\ref{eq:u}) following the same method as
the one employed above to calculate $D_{2n+1}$.

\section{The case of the rectangular distribution}
It is also instructive to consider the case of a rectangular
distribution of random fields (i.e.
$\rho(x)= 1/2\Delta \quad { \textrm{if}\quad  \Delta  <  x } $ 
and  zero otherwise), since all the calculations can be
carried out explicitely. 
As usual, we cycle the field around $H=0$ and we take $H_0=J$. 
The calculation should be divided in several cases, 
depending on the value of $\Delta$.

(i) For $\Delta \ge 3J$,  we have
$\rho(x)=\rho(2J-x)=1/2\Delta$, so that 
$p_k(H^*)-p_k(-H^*)=H^*/\Delta$ and Eq.~(\ref{eq:u*}),
reduces to 
\begin{equation}\label{eq:u*2}
U^2(H^*)=\frac{1}{4}\left(\frac{\Delta-J}{\Delta-H^*}\right)^2
\end{equation}
Inserting these results in Eq.~(\ref{mm1}), we obtain:
\begin{equation}
M(H^*)=\left(\frac{\Delta-J}{\Delta-H^*}\right)^2\frac{H}{\Delta}.
\label{eq:munif}
\end{equation}
Expanding Eq.~\ref{eq:munif}, we obtain the values for $a$ and  $b$ 
\begin{equation}\label{abuni}
\left\{\begin{array}{l}
a=\frac{1}{\Delta}\left[1-\frac{J}{\Delta}\right]^2   \\
\\ 
b=2\frac{1}{\Delta^2}\left[1-\frac{J}{\Delta}\right]^2
\end{array}
\right.
\end{equation}

(ii) For $ 2J<\Delta < 3J$,  $U^2(0)$ is still given by
Eq.~(\ref{eq:u*2}) but $p_k(H^*)$ differs from the previous case.
The magnetization is now given by
\begin{equation}\label{eq:m2}
\left\{\begin{array}{l}
M(H^*)=\left(\frac{\Delta-J}{(\Delta-H^*)}\right)^2\frac{3H^*-2J+\Delta}{4\Delta}
~~~~H^*>\Delta-2J\\
M(H^*)=\left(\frac{\Delta-J}{(\Delta-H^*)}\right)^2\frac{H}{\Delta}~~~
H^*<\Delta-2J.
\end{array}
\right.
\end{equation}
The expansion around $H^*$ is thus still given by Eq.~(\ref{abuni}).

(iii) The behavior for $J<\Delta<2J$ is again different:
close to $H^*=0$ the peak magnetization is not given by Eq.~(\ref{mm1}),
but for $H^*<2J-\Delta$ can be written as
\begin{equation}
M(H^*)=\frac{(\Delta-J)^2H^*}{4J\Delta(\Delta-H^*)},
\end{equation}
so that expanding we obtain
 \begin{equation}
\left\{\begin{array}{l}
a=\frac{(\Delta-J)^2}{4\Delta^2J}  
\\ 
b=\frac{(\Delta-J)^2}{2\Delta^3J}
\end{array}
\right.
\end{equation}

(iv) Finally for $\Delta<J$ there is no hysteresis and thus the
Rayleigh law is not defined.

\begin{thebibliography}{10}

\bibitem{RAY-87}
L. Rayleigh, Philos. Mag., Suppl. {\bf 23},  225  (1887).

\bibitem{Bertotti}
G. Bertotti, {\em Hysteresis in Magnetism} (Academic Press, San Diego, 1998).

\bibitem{DAM-97}
D. Damjanovic, J. Appl. Phys. {\bf 82},  1788  (1997). For a review see Rep. Prog. Phys.
{\bf 61}, 1267 (1998).

\bibitem{BOL-00}
D. Bolten et al., Appl. Phys. Lett. {\bf 77},  3830  (2000).

\bibitem{NEE-42}
L. N{\'e}el, Cah. Phys. {\bf 12},  1  (1942).

\bibitem{PFE-67}
K.~H. Pfeffer, Phys. Stat. Sol. {\bf 21},  857  (1967).

\bibitem{VER-74}
R. Vergne, Z. Blazek and J. L. Porteseil, 
Phys. Stat. Sol. A  {\bf 25}, 171 (1974).

\bibitem{KRO-92}
H. Kronm\"uller and T. Reininger, J. Magn. Magn. Mat. {\bf 112},  1  (1992).

\bibitem{MAG-99}
A. Magni, C. Beatrice, G. Durin, and G. Bertotti, J. Appl. Phys. {\bf 86},
  3253  (1999).

\bibitem{HER-97}
G. Herzer,  in {\em Handbook of Magnetic Materials, Vol. 10}, edited by
  K.~H.~J. Buschow (Elsevier, Amsterdam, 1997), p.\ 415.

\bibitem{SUZ-91}
K. Suzuki, A. Makino, A. Inoue, and T. Masumoto, J. Appl. Phys {\bf 70},  6232
  (1991).

\bibitem{LEU-97}
M.~S. Leu and T.~S. Chin, J. Appl. Phys {\bf 81},  4051  (1997).

\bibitem{LIM-93}
S.~H. Lim et al.,  J. Appl. Phys {\bf
  73},  6591  (1993).

\bibitem{SET-93}
J.~P. Sethna et. al, Phys. Rev.
  Lett. {\bf 70},  3347  (1993)

\bibitem{DAH-96}
K. Dahmen and J.~P. Sethna, Phys. Rev. B {\bf 53},  14872  (1996).

\bibitem{PER-99}
O. Perkovic, K.~A. Dahmen, and J.~P. Sethna, Phys. Rev. B {\bf 59},  6106
  (1999).

\bibitem{VIV-94}
E. Vives and A. Planes, Phys. Rev. B {\bf 50},  3839  (1994);ibid. 
{\bf 63},  134431  (2001).

\bibitem{DAS-99}
R. da~Silveira and M. Kardar, Phys. Rev. E {\bf 59},  1355  (1999).

\bibitem{BER-00}
A. Berger et al., Phys. Rev.
  Lett. {\bf 85},  4176  (2000).

\bibitem{SHU-96}
P. Shukla, Physica A {\bf 233},  235  (1996).

\bibitem{SHU-00}
P. Shukla, Phys. Rev. E {\bf 62},  4725  (2000).

\bibitem{DHA-97}
D. Dhar, P. Shukla, and J.~P. Sethna, J. Phys. A {\bf 30},  5259  (1997).

\bibitem{SHU-01}
P. Shukla, Phys. Rev. E {\bf 63},  027102  (2001).

\bibitem{CAR-01}
J.~H. Carpenter et al. , J. Appl. Phys. {\bf 89},  6799  (2001).

\bibitem{BER-90}
G. Bertotti and M. Pasquale, J. Appl. Phys. {\bf 67}, 5255 (1990).

\bibitem{nota1} Notice that these equations are only valid for
$|H_n| \le H_0-2J$.

\bibitem{nota2} For instance the limit can be taken chosing
$H_{n}=(-1)^n(1-\epsilon)^n J$ and then expanding all the
quantities for $\epsilon \to 0^+$. 


\bibitem{COL-01}
F. Colaiori, A. Gabrielli and S. Zapperi,
preprint, cond-mat/0112190.

\bibitem{ALA-01}
M. Alava, P. Duxbury, C. Moukarzel, and H. Rieger,  in {\em Phase transitions
  and critical phenomena, Vol 18}, edited by C. Domb and J. Lebowitz (Academic
  Press, San Diego, 2001).

\bibitem{ZAR-01}
A recent investigation of such an ``hysteretic optimization'' 
can be found in G. Zarand, F. Pazmandi, K.F. Pal and G.T. Zimanyi,
preprint, cond-mat/0109359.

\end{thebibliography}
\end{document}